\documentclass[a4paper,11pt]{article}
\usepackage{mms}
\pdfoutput=1 
\usepackage[T1]{fontenc} 

\def\be{\begin{equation}}
\def\ee{\end{equation}}
\def\bea{\begin{eqnarray}}
\def\eea{\end{eqnarray}}

\def\d#1#2{\frac{\displaystyle #1}{\displaystyle #2}}
\def\no{\nonumber}

\def\La{\Lambda}
\def\m{\mu}
\def\n{\nu}

\begin{document}


\title{\boldmath $(2+1)$-dimensional regular black holes with nonlinear electrodynamics sources}

\author[b]{Yun He,}
\author[a,b]{Meng-Sen Ma\footnote{Corresponding author. E-mail address: mengsenma@gmail.com}}


\affiliation[a]{Institute of Theoretical Physics, Shanxi Datong
University, Datong 037009, China}
\affiliation[b]{Department of Physics, Shanxi Datong
University, Datong 037009, China}









\abstract{On the basis of two requirements: the avoidance of the curvature singularity and the Maxwell theory as the weak field limit of the nonlinear electrodynamics, we find two restricted conditions on the metric function of $(2+1)$-dimensional regular black hole in general relativity coupled with nonlinear electrodynamics sources. By the use of the two conditions, we obtain a general approach to construct $(2+1)$-dimensional regular black holes. In this manner, we construct four $(2+1)$-dimensional regular black holes as examples. We also study the thermodynamic properties of the regular black holes and verify the first law of black hole thermodynamics.  }

\keywords{regular black hole, ~nonlinear electrodynamics, ~$(2+1)$-dimensional spacetime}

\maketitle
\flushbottom


\section{Introduction}

It has long been known that there is a large class of regular black holes,  for which spacetime singularities can be
avoided. The first example of a regular black hole was constructed by Bardeen in 1968\cite{Bardeen}.
Nearly 30 years later, Ay\'{o}n-Beato et al. reobtained the Bardeen black hole by describing it as the gravitational field of a
kind of nonlinear magnetic monopole\cite{AyonBeato.149.2000}. Regular black holes can be constructed in different circumstances.
Dymnikova had ever proposed a kind of nonsingular black holes, with a de Sitter core smoothly connecting to a Schwarzschild outer geometry\cite{Dymnikova.235.1992}.
The noncommutative geometry inspired black hole is also a kind of regular black hole with the naked singularity replaced by a de Sitter, regular geometry around
the origin\cite{Nicolini.547.2006}.
More regular black holes can be constructed by introducing nonlinear electromagnetic sources or scalar field\cite{AyonBeato.5056.1998,AyonBeato.25.1999,AyonBeato.629.1999,Bronnikov.044005.2001,Burinskii.104017.2002,Dymnikova.4417.2004,Bronnikov.251101.2006,Ma.529.2015,Balart.14.2014,Balart.124045.2014,Fan.124027.2016}.
For details of regular black holes, one can refer to the papers \cite{Lemos.124005.2011, Ansoldi.0802.2008} and and references therein.

$(2+1)$-dimensional gravity is usually studied as a toy model with the hope that it can shed light on some problems of its $(3+1)$-dimensional counterpart.
In $(2+1)$-dimensional spacetime, general relativity does not reduce to the Newtonian theory in the weak field limit and gravitational field has no dynamical degrees of freedom \cite{Gott.243.1984,Barrow.551.1986,Gott.1019.1986}. It is generally believed that there is no black hole in $(2+1)$-dimensional gravity\cite{Giddings.751.1984}. Not surprisingly, the discovery of the famous BTZ black hole\cite{Banados.1849.1992,Banados.1506.1993} aroused much interest in the study on $(2+1)$-dimensional black holes\cite{Chan.6385.1994,Zaslavskii.L33.1994,Carlip.2853.1995,Kamata.196.1995,Lemos.46.1995,Cataldo.2971.1996,Clement.70.1996,Hirschmann.5579.1996,Martinez.104013.2000,Yamazaki.2001,Mazharimousavi.124021.2011,Gurtug.104004.2012,Xu.124008.2013,Frassino.124069.2015}.

In lower dimensional spacetime, there are also regular black holes. A nonsingular $(1+1)$-dimensional black hole was derived by using a non-dynamical scalar field\cite{Trodden.483.1993}.
In \cite{Cataldo.084003.2000}, a $(2+1)$-dimensional regular black hole was constructed by introducing the nonlinear electrodynamic source. Myung et al. found a regular black hole in $(2+1)$-dimensional anti-de Sitter space by introducing an anisotropic perfect fluid inspired by the noncommutative black hole\cite{Myung.405.2009}. By employing nonlinear Born-Infeld electrodynamics with the Hoffmann term and gluing different spacetimes, a regular extension of the charged BTZ black hole was constructed in \cite{Mazharimousavi.893.2012}.

In this letter, we propose a procedure by which one can construct many $(2+1)$-dimensional regular black holes in general relativity coupled
to nonlinear electrodynamics (NED). The conventional approach to derive the regular black hole is to start from the energy-momentum tensor or the Lagrangians of the given matter fields, and then solve the field equation. Here we go the opposite way. As the authors did in \cite{Hayward.031103.2006,Balart.14.2014,Balart.124045.2014,Fan.124027.2016}, we can first construct the regular black holes according to the requirement of avoidance of the curvature singularity and then derive the corresponding energy-momentum or the matter fields. As examples, we construct four $(2+1)$-dimensional regular black holes and study their thermodynamic properties.

The rest of the paper is arranged as follows.
In section \ref{review}, we provide a brief review of $(2+1)$-Einstein theory with nonlinear electrodynamics.
In section \ref{3dBH}, we first analyze the conditions under which the black hole solution is free of curvature singularity and then construct several regular black holes.
In sections \ref{thermo}, we calculate the thermodynamic quantities of these regular black holes and verify the first law of black hole thermodynamics.
 The conclusion is given in section \ref{conclusion}. In this paper we set $(G=c=\hbar=k_B=1)$.

\section{$(2+1)$-Einstein theory with nonlinear electrodynamics
\label{review}}

The action of the $(2+1)$-Einstein gravity coupled with nonlinear electrodynamics is given by
\be\label{action}
S=\int d^3x\sqrt{-g}\left[\frac{R-2\Lambda}{16\pi}+L(F)\right],
\ee
where $g$ is the determinant of the metric tensor, $\Lambda=-1/l^2$ is the cosmological constant, and $L(F)$ is the Lagrangian of the nonlinear electrodynamics with $F=F^{\mu\nu}F_{\mu\nu}$.

Variation with respect to the metric tensor, one can derive the field equation:
\be\label{GReq}
G_{\mu\nu}+\Lambda g_{\mu\nu}=8\pi T_{\mu\nu},
\ee
with
\be
T_{\mu\nu} \equiv -\d{2}{\sqrt{-g}}\d{\delta(\sqrt{-g} L_m)}{\delta g^{\mu\nu}}=g_{\mu\nu}L(F)-4L_{,F}F_{\mu\alpha}F_{\nu}^{~\alpha},
\ee
where $L_{,F}$ represents the derivative of $L(F)$ with respect to $F$. The electromagnetic field equation is
\be\label{EMeq}
\nabla_\mu(L_{,F}F^{\mu\nu})=0.
\ee
For the static, circularly symmetric spacetime we take the simplest metric ansatz
\be\label{metric}
ds^2=-f(r)dt^2+f(r)^{-1}dr^2+r^2d\phi^2.
\ee
Under this metric ansatz, it has been shown in \cite{Cataldo.084003.2000} that the magnetic field vanishes and only the electric field plays the role of the source of gravitational field. Therefore, the electromagnetic field tensor takes the form of
 \be
 F_{\mu\nu}=E(r)(\delta^t_{\m}\delta^r_{\n}- \delta^t_{\n}\delta^r_{\m}),
 \ee
 and correspondingly $F=-2E^2$.

 The field equations under the metric ansatz have the simple forms:
\be\label{twoeqs}
\d{f'(r)}{2r}+\La=8\pi\left[L(F)+4L_{,F}E(r)^2\right], \quad E(r)L_{,F}=-\d{q}{r},
\ee
or
\be\label{twoeqs2}
\d{f'(r)}{2r}+\La=8\pi\left[L(r)-\frac{4q}{r}E(r)\right], \quad L'(r)=\d{4q}{r}E'(r),
\ee
where $q$ is an integration constant related the the electric charge. Clearly, in the two equations there are three unknown function: $f(r),~L(F),~E(r)$. Generally, $L(F)$ is first given and then one can solve the equations to obtain $f(r)$ and $E(r)$. However, in this way the obtained black hole solutions are  not regular in general. So, in this paper we take the opposite tack. We first construct the regular black holes and then derive the concrete forms of the Lagrangian of the nonlinear electrodynamics according to the field equations.

\section{$(2+1)$-dimensional regular black holes}
\label{3dBH}

The black holes are regular if the curvature invariants $R,~R_{\mu\nu}R^{\mu\nu}$ are regular at $r=0$\footnote{The `` regular '' in this paper means the avoidance of the curvature singularity.  In $2+1$ spacetime, there may be conical singularity at $r=0$\cite{Gott.243.1984}. One can, in fact, eliminate the conical singularity by choosing the proper integration constants\cite{Cataldo.084003.2000}.}. For the metric ansatz we considered, these invariants are
\bea
R&=&-f''(r)-\frac{2 f'(r)}{r}, \no \\
R_{\mu\nu}R^{\mu\nu}&=&\frac{1}{2} f''(r)^2+\frac{3 f'(r)^2}{2 r^2}+\frac{f'(r) f''(r)}{r}.
\eea
It can be easily found that if
\be
\lim_{r\rightarrow 0}\frac{f'(r)}{r}=\text{\emph{constant}},
\ee
the above curvature invariants are all regular and finite in the limit $r \rightarrow 0$\footnote{When this limit is established, naturally there is $\lim_{r\rightarrow 0}f''(r)=const$ due to the L'Hospital's rule.}.
This is a sufficient but not necessary condition to judge the regularity of a black hole solution. By this standard, we find that the simplest regular black hole in $(2+1)$-dimensional spacetime is $f(r)=C+D r^2$, which is in fact the static BTZ black hole with $f(r)=-m + r^2/l^2$. When matter fields exist, the metric function may include other terms. So we conjecture that the metric function of general regular black holes in $(2+1)$-dimensional spacetime should takes the form
\be\label{kr}
f(r)=-m +\d{r^2}{l^2}+k(r),
\ee
where $k(r)$ is at least a two times differentiable function and can be expanded in such a form of series:
\be\label{k1}
k(r)=k_0+k_2 r^2+O(r^3).
\ee
$k_0,~k_2$ are two constants. There are many functions which have the above asymptotic behavior. In this way, through choosing proper $k(r)$ one can construct various regular black holes.

One thing that should be noted is that in the above procedure of constructing the regular black holes we only require the finite curvature invariants at $r=0$ and require nothing on the Lagrangian of the nonlinear electrodynamics.  To obtain more physical results, we need to add two requirements or conditions on the nonlinear electrodynamics, which are: (1) the nonlinear electrodynamics should reduce to the Maxwell theory in the weak field limit, namely $L(F)\rightarrow F$, for large $r$ and (2) the weak energy condition should be fulfilled.

The condition (1) will put more constraints on the function $k(r)$ in Eq.(\ref{kr}). Considering $L(F)\rightarrow F$, the electric field has the asymptotic solution $E(r) \sim \d{1}{r}$.
Substituting them into Eq.(\ref{twoeqs}), we find that for large $r$
\be\label{k2}
k'(r) \sim \d{1}{r}.
\ee
Therefore, to construct a more physically acceptable regular black hole, the function $k(r)$ should both satisfy Eq.(\ref{k1}) and Eq.(\ref{k2}).

The weak energy condition ensures that an observer measures a non-negative energy density. For our nonlinear electrodynamics, it means
\be\label{WEC}
L+4E^2L_{,F}\leq 0.
\ee
This condition can be easily fulfilled by many regular black holes.

Below we list four regular black holes constructed according to the approach above.

\textbf{Case I}:
\be
f(r)=-m+\d{r^2}{l^2} -q^2\sqrt{a^2+r^2}, \no
\ee
and
 \be
  L(r)=-\frac{a^2 q^2}{16 \pi  \left(a^2+r^2\right)^{3/2}}, \quad E(r)=\frac{q r^3}{64 \pi  \left(a^2+r^2\right)^{3/2}}.
  \ee
 Considering the length dimension, in this case $[a]=1$ and $[q]=-1$.  This regular black hole does not satisfy the Maxwell limit in the weak field approximation because $E(r) \rightarrow const $ for very large $r$. Besides, the thermodynamic quantities of the black hole is problematic. For this simple example, we just want to show that one can construct many, even infinite many this kind of regular black holes if only requiring to be free of curvature singularity. In this paper we are not concerned with this type.

\bigskip

\textbf{Case II}:
\be
f(r)=-m+\d{r^2}{l^2} -q^2 \ln \left(\d{q^2/a^2+r^2}{l^2}\right), \no
\ee
and
 \be
  L(r)=\frac{q^2 \left(a^4r^2-a^2q^2\right)}{8 \pi  \left(q^2+a^2r^2\right)^2}, \quad E(r)=\frac{a^4q r^3}{16 \pi  \left(q^2+a^2r^2\right)^2}.
  \ee
 This is just the regular black hole obtained previously by Cataldo and Garc\'{\i}a in \cite{Cataldo.084003.2000}, but with two differences. In \cite{Cataldo.084003.2000}, the logarithmic term is $q^2\ln (r^2+a^2)$. Here we divide $l^2$ in the logarithmic term to ensure the dimensionless result. The inclusion of the $l^2$ in the logarithmic term does not influence the electric field.  Besides, we replace the $a^2$ in the logarithmic term with $\d{q^2}{a^2}$. This modification will affect the electric field and the Lagrangian of nonlinear electrodynamics, correspondingly.

\bigskip

\textbf{Case III}:

\be\label{C3}
f(r)=-m+\d{r^2}{l^2}-2q^2\left[\ln \left(\frac{q}{a l}+\frac{r}{l}\right)+\frac{q}{a r+q}\right], \no
\ee
and
 \be
  L(r)=\frac{a^2 q^2 (a r-q)}{8 \pi  (a r+q)^3}, \quad E(r)=\frac{a^3 q r^2}{16 \pi  (a r+q)^3}.
  \ee

\bigskip

\textbf{Case IV}:
\be\label{C4}
f(r)=-m+\d{r^2}{l^2} -q^2 \sinh ^{-1}\left(\d{a^2 r^2}{q^2}\right), \no
\ee
and
 \be
  L(r)=\frac{a^2 q^2 \left(a^4 r^4-q^4\right)}{8 \pi  \left(a^4 r^4+q^4\right)^{3/2}}, \quad E(r)=\frac{a^6 q r^5}{16 \pi  \left(a^4 r^4+q^4\right)^{3/2}}.
  \ee

In the Cases II-IV, the length dimensions of the parameters ($a$, $q$) are $[q^2]=-1$ and $[a^2]=-3$. The nonlinear electrodynamics in the Cases II-IV all have the Maxwell theory as the limit in the weak field approximation. Clearly, when $a \rightarrow \infty$ or $r \rightarrow \infty$, the electric field tends to $q/r$. In particular, for the Case II and Case III, the regular black holes will reduce to the static, charged BTZ black hole when the parameter $a \rightarrow \infty$.

In the four examples, we obtain $L(r)$ and $E(r)$ as functions of $r$. In principle, one can always combine them to give $L(F)$. One can easily check that the electromagnetic fields in the four cases all satisfy the weak energy condition.

\section{Thermodynamics of the regular black holes}
\label{thermo}

The properties of the regular black hole in Case II have been studied by Cataldo and Garc\'{\i}a\cite{Cataldo.084003.2000}.
In this section, we are concerned with the regular black holes in Case III and Case IV. Below we only consider the cases with positive $(m,~q)$.

\begin{figure}[htp]
\center{
\includegraphics[width=5cm,keepaspectratio]{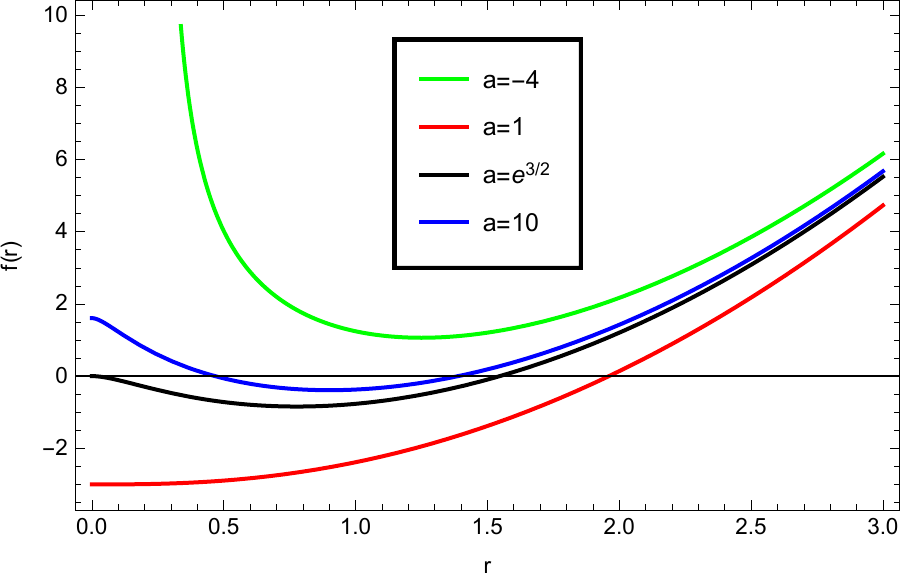} \hspace{0.5cm}
\includegraphics[width=5cm,keepaspectratio]{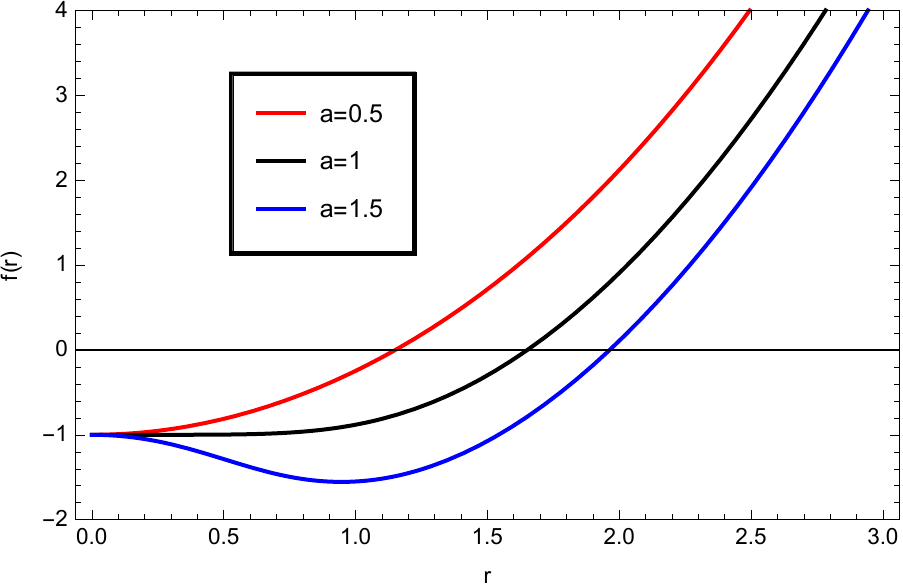}
\caption{Behaviors of the metric function $f(r)$ for different values of $a$. The left panel corresponds to the regular black hole in Case III and the right panel corresponds to the one in Case IV. We set $m=l=q=1$. }\label{fr34}}
\end{figure}

For event horizon to exist, the metric function $f(r)$ must have zeros. The situation is illustrated in Fig.\ref{fr34}. For the regular black hole in Case IV, $f(r)=0$ always has one and only one solution, which is the event horizon of the regular black hole. For the regular black hole in Case III, $f(r)=0$ can have zero, one or two solutions for different values of the parameter $a$. When $a$ is negative, no horizon exists. When $0<a<e^{3/2}$, there is only one horizon and when $a> e^{3/2}$, no matter how large, there  are always two horizons.

\begin{figure}[htp]
\center{
\includegraphics[width=5cm,keepaspectratio]{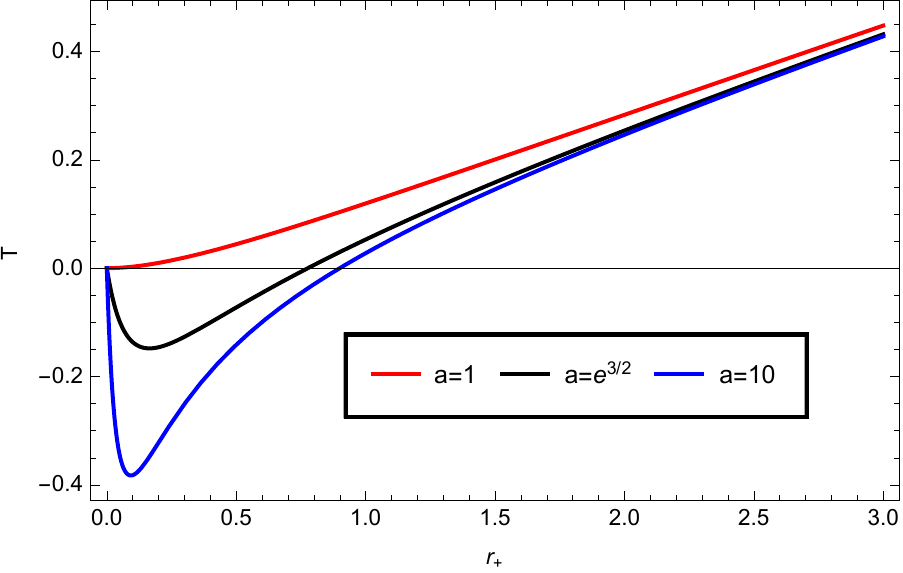}\hspace{0.5cm}
\includegraphics[width=5cm,keepaspectratio]{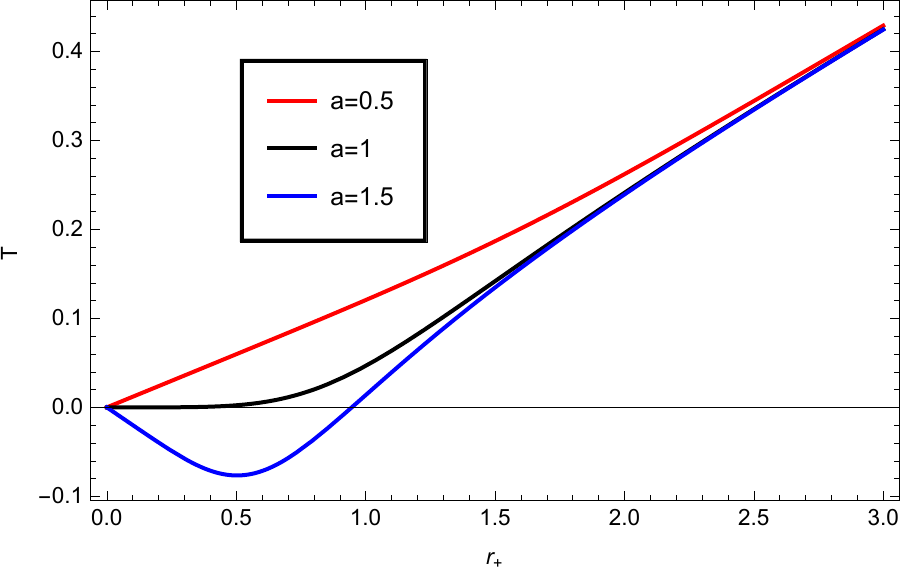}
\caption{The temperature of the regular black holes as functions of $r_{+}$. The left panel corresponds to the regular black hole in Case III and the right panel corresponds to the one in Case IV. We set $l=q=1$. }\label{T34}}
\end{figure}

The temperatures of the two regular black holes can be calculated directly. In Case III, it is
\be\label{T3}
T=\d{f'(r_{+})}{4\pi}=\frac{r_{+} }{2 \pi  l^2 }\left[1-\d{a^2l^2q^2}{(a r_{+}+q)^2}\right],
\ee
and in Case IV it is
\be\label{T4}
T=\frac{r_{+} }{2 \pi  l^2 }\left(1-\frac{a^2 q^2l^2}{\sqrt{a^4 r_{+}^4+q^4}}\right),
\ee
where $r_{+}$ represents the position of the event horizon. Comparing Eq.(\ref{T3}) with Eq.(\ref{T4}), the temperatures in the two cases have similar form and both asymptotically approach the temperature of static, charged BTZ black hole.
As is depicted in Fig.\ref{T34}, when $a$ is small enough the temperatures can be always positive and increases monotonically with the increase of $r_{+}$. For larger $a$, the temperatures are negative for small $r_{+}$  and are positive for large $r_{+}$.

The entropy of the regular black holes in $2+1$ dimensions takes the general Bekenstein-Hawking form: $S=A/4=\pi r_{+}/2$.

There are two parameters: $a$ and $q$, in the solutions listed in Case III and Case IV.  We should judge which parameters are black hole parameters(integration constants) and which are model parameters(coupling constants). In fact,  the  $L(r)$ and $E(r)$ in Case III and Case IV can be respectively rewritten as
 \be
  L(r)=\frac{r/q-1/a}{8 \pi  (r/q+1/a)^3}, \quad E(r)=\frac{r^2/q^2}{16 \pi  (r/q+1/a)^3}.
  \ee
 \be
  L(r)=\frac{ r^4/q^4-1/a^4}{8 \pi  \left(r^4/q^4+1/a^4\right)^{3/2}}, \quad E(r)=\frac{r^5/q^5}{16 \pi  \left(r^4/q^4+1/a^4\right)^{3/2}}.
  \ee
Clearly, $r/q$ can be eliminated from $L(r)$ and $E(r)$. Then $L(E)$ only contains the parameter $a$, which is a coupling constant. For our solutions there are two black hole parameters: $m$ and $q$, which should be related to the black hole mass $M$ and electric charge $Q$, respectively.


The electric charge can also be derived through comparing with the charged BTZ black hole. However, one can also directly employ the formula given in \cite{Miskovic.024011.2011}: $Q=4\text{Vol}(\Gamma_{D-2})q$, 
where $\text{Vol}(\Gamma_{D-2})$ is the unit volume of a $D-2$-dimensional Riemann space. In $(2+1)$-dimensional case, we obtain $Q=8\pi q$.

Comparing with the static, charged BTZ black hole, one can easily obtain the mass of the regular black holes $M=m/8$. We can further confirm this result according to the first law of the black hole thermodynamics: $dM=TdS+\Phi dQ$. Form the first law, there should be $T=\left.\d{\partial M}{\partial S}\right|_{Q}$.
On the other hand, our metric functions all have the form: $f(r)=-m +\d{r^2}{l^2}+k(r)$, which means 
\be
T=\d{f'(r_{+})}{4\pi}=\d{1}{4\pi}\left.\d{\partial m}{\partial r_{+}}\right|_{q}=\d{1}{8}\left.\d{\partial m}{\partial S}\right|_{q}=\d{1}{8}\left.\d{\partial m}{\partial S}\right|_{Q}.
\ee
Compared with the first law, we know that the black hole mass $M=m/8$.

$\Phi$ is the electric potential at infinity measured with respect to the event horizon. In general,
it is defined by $\Phi=\phi(\infty)-\phi(r_{+})$ with $\phi(r)$ satisfying $\phi(r)=\int_{r}^{\infty} E(x)dx$.

In $(3+1)$- or higher dimensional spacetime, one can directly derive the $\Phi$ through the integral of the electric field.
Whereas, in the $(2+1)$-dimensional spacetime one cannot obtain $\Phi$ in this way. As mentioned in section \ref{3dBH}, if the nonlinear electrodynamics reduces to the Maxwell theory in the weak field limit, the electric field $E(r) \sim 1/r$ for large $r$.
Thus, the integral of the electric field  will give a logarithmic divergent result at infinity.
However, if $\Phi$ has been derived in other methods, one can verify it with the help of  the relation $E(r_{+})=\nabla \Phi(r_{+})$.

For the regular black holes in Case III and Case IV, the black hole masses are respectively
\bea
M&=& \d{r_{+}^2}{8l^2}- \d{q^2}{4} \left[\ln \left(\frac{q}{a l}+\frac{r_{+}}{l}\right)+\frac{q}{a r_{+}+q}\right], \no \\
M&=&\d{r_{+}^2}{8l^2}-\d{q^2}{8} \left[\sinh ^{-1}\left(\frac{a^2 r_{+}^2}{q^2}\right)\right].
\eea
According to the first law, the electric potentials in the two cases should be
\bea
\Phi&=&\left.\d{\partial M}{\partial Q}\right|_{S}=\frac{q^2 (4 a r_{+}+3 q)}{32 \pi  (a r_{+}+q)^2}+ \d{q}{16\pi} \ln \left(\d{a r_{+}+q}{al}\right), \no \\
\Phi&=&\d{q}{32\pi}\left[\sinh ^{-1}\left(\frac{a^2 r_{+}^2}{q^2}\right)-\frac{a^2 r_{+}^2}{\sqrt{a^4 r_{+}^4+q^4}}\right].
\eea
Taking the derivative with respect to $r_{+}$, we find that they indeed give the same electric fields as those in Eq.(\ref{C3}) and Eq.(\ref{C4}).

Thus, the two $(2+1)$-dimensional regular black holes indeed satisfy the first law of black hole thermodynamics. This result is what we expect. After all,
the first law, Smarr formula and the Komar charge may be violated for some regular black holes in $(3+1)$-dimensional spacetime\cite{Breton.643.2005,Balart.280.2010,Ma.245014.2014,Ma.CPL}.
In fact, the two $(2+1)$-dimensional regular black holes also violate the Smarr formula. We can remedy this problem by slightly adjusting the $k(r)$ part in the metric functions and treating the cosmological constant as the thermodynamic pressure, $P=-\Lambda/8\pi=1/8\pi l^2$. In this extended phase space, the Smarr formula can be established. Moreover, the first law in the extended phase space is also fulfilled. The only drawback is that the parameter $l$ will occur in the electric field. Some details of this problem are presented in the Appendix \ref{A}.

At last, we discuss the thermodynamic stability of the regular black holes. According to the first law, one can define the heat capacity as $C=\left.\d{\partial M}{\partial T}\right|_{Q}=\left.T\d{\partial S}{\partial T}\right|_{Q}$. As is shown in Fig.\ref{T34}, the positive temperatures increase monotonically with the increase of $r_{+}$, namely $S$. Thus, the heat capacity is always positive. Therefore, the two regular black holes are always thermodynamically stable.

\section{Conclusion \label{conclusion}}

In this paper, we studied the $(2+1)$-dimensional regular black holes in the general relativity coupled to nonlinear electrodynamics.
Under the metric ansatz Eq.(\ref{metric}), we analyzed the curvature invariants in $(2+1)$-dimensional spacetime. For such a metric function, $f(r)=-m +r^2/l^2+k(r)$, we found that so long as $k(r)=k_0+k_2 r^2+O(r^3)$, the black holes will be regular. According to this standard, one can construct many $(2+1)$-dimensional regular black holes. If further requiring that the nonlinear electrodynamics can reduce to the Maxwell theory in the weak field limit, the metric function will receive another constraint: $k'(r) \sim \d{1}{r}$ for large $r$. On the basis of these constraints, we construct four regular black holes as examples and derived the corresponding electric fields and the Lagrangians of the nonlinear electrodynamics.

We then analyzed the thermodynamic properties of the regular black holes in the Case III and Case IV.
We found that their temperatures have the similar form and both asymptotically approach that of the static, charged BTZ black hole. When the parameter $a$ is small enough, the temperatures can be always positive. We also derive other thermodynamic quantities, such as the black hole mass, electric charge and the electric potential. For the two regular black hole, the first law of black hole thermodynamics is  completely fulfilled. The positive temperatures are monotonic functions of the black hole entropy. This means that the heat capacity is always positive and thus the black holes are thermodynamically stable.


\appendix
\section{Regular black holes in extended phase space}
\label{A}

To satisfy both the first law of black hole thermodynamics and the Smarr formula, the regular black holes constructed above should be modified slightly.

Taking the Case II as an example, the regular black hole satisfies the first law: $dM=TdS+\Phi dQ$, but does not satisfy the Smarr formula. While, for the slightly modified regular black hole
\be
f(r)=-m+\d{r^2}{l^2} -q^2 \ln \left(\d{q^2}{a^2}+\d{r^2}{l^2}\right),
\ee
after considering  $P=-\Lambda/8\pi=1/8\pi l^2$, the first law
\be
dM=TdS+\Phi dQ +VdP,
\ee
and the Smarr formula $TS-2VP=0$ are both fulfilled in the extended phase space. Here $V$ is not the standard thermodynamic volume $\pi r_{+}^2$, but
\be
V=\left.\d{\partial M}{\partial P}\right|_{S,Q}=\pi r_{+}^2\left(1-\d{q^2a^2l^2}{q^2l^2+a^2r_{+}^2}\right),
\ee
which violates the isoperimetric inequality\cite{Frassino.124069.2015,Cvetic.2011}.

The drawback of this modification is that the de Sitter radius $l$ will occur in the electric field and the Lagrangian of the NED:
\be
  L(r)=\frac{q^2 \left(a^4r^2-a^2q^2l^2\right)}{8 \pi  \left(q^2l^2+a^2r^2\right)^2}, \quad E(r)=\frac{a^4q r^3}{16 \pi  \left(q^2l^2+a^2r^2\right)^2}.
 \ee

%
%

\acknowledgments

This work is supported in part by the National Natural Science Foundation
of China (Grant Nos.11605107) and by the Natural Science Foundation of Shanxi (Grant No.201601D021022).


\bibliographystyle{JHEP}

%
%
%
%
%
%
%
%
%
%


\begin{thebibliography}{99}



\bibitem{Bardeen} J. M. Bardeen, in Conference Proceedings of GR5, Tbilisi, USSR, 1968, p. 174.

\bibitem{AyonBeato.149.2000}
E.~Ay{\'o}n-Beato and A.~Garc\'{\i}a, \emph{{The Bardeen model as a nonlinear
  magnetic monopole}},
  \href{http://dx.doi.org/10.1016/S0370-2693(00)01125-4}{\emph{Phys. Lett. B}
  {\bf 493} (2000) 149--152}.

\bibitem{Dymnikova.235.1992}
I.~Dymnikova, \emph{{Vacuum nonsingular black hole}},
  \href{http://dx.doi.org/10.1007/BF00760226}{\emph{Gen. Relativ. Gravit.} {\bf
  24} (1992) 235--242}.

\bibitem{Nicolini.547.2006}
P.~Nicolini, A.~Smailagic and E.~Spallucci, \emph{{Noncommutative geometry
  inspired Schwarzschild black hole}},
  \href{http://dx.doi.org/10.1016/j.physletb.2005.11.004}{\emph{Phys. Lett. B}
  {\bf 632} (2006) 547--551}.


%

\bibitem{AyonBeato.5056.1998}
E.~Ay{\'o}n-Beato and A.~Garc\'{\i}a, \emph{{Regular Black Hole in General
  Relativity Coupled to Nonlinear Electrodynamics}},
  \href{http://dx.doi.org/10.1103/PhysRevLett.80.5056}{\emph{Phys. Rev. Lett.}
  {\bf 80} (1998) 5056--5059}.

\bibitem{AyonBeato.25.1999}
E.~Ay{\'o}n-Beato and A.~Garc\'{\i}a, \emph{{New regular black hole solution
  from nonlinear electrodynamics}},
  \href{http://dx.doi.org/10.1016/S0370-2693(99)01038-2}{\emph{Phys. Lett. B}
  {\bf 464} (1999) 25--29}.

\bibitem{AyonBeato.629.1999}
E.~Ayon-Beato, \emph{{Non-Singular Charged Black Hole Solution for Non-Linear
  Source}}, \href{http://dx.doi.org/10.1023/A:1026640911319}{\emph{Gen.
  Relativ. Gravit.} {\bf 31} (1999) 629--633}.

\bibitem{Bronnikov.044005.2001}
K.~A. Bronnikov, \emph{{Regular magnetic black holes and monopoles from
  nonlinear electrodynamics}},
  \href{http://dx.doi.org/10.1103/PhysRevD.63.044005}{\emph{Phys. Rev. D} {\bf
  63} (2001) 044005}.

\bibitem{Burinskii.104017.2002}
A.~Burinskii and S.~R. Hildebrandt, \emph{{New type of regular black holes and
  particlelike solutions from nonlinear electrodynamics}},
  \href{http://dx.doi.org/10.1103/PhysRevD.65.104017}{\emph{Phys. Rev. D} {\bf
  65} (2002) 104017}.

\bibitem{Dymnikova.4417.2004}
I.~Dymnikova, \emph{{Regular electrically charged vacuum structures with de
  Sitter centre in nonlinear electrodynamics coupled to general relativity}},
  \href{http://dx.doi.org/10.1088/0264-9381/21/18/009}{\emph{Class. Quantum
  Grav.} {\bf 21} (2004) 4417--4428}.

\bibitem{Bronnikov.251101.2006}
K.~A. Bronnikov and J.~C. Fabris, \emph{{Regular phantom black holes}},
  \href{http://dx.doi.org/10.1103/PhysRevLett.96.251101}{\emph{Phys. Rev.
  Lett.} {\bf 96} (2006) 251101}.


\bibitem{Ma.529.2015}
M.-S. Ma, \emph{{Magnetically charged regular black hole in a model of
  nonlinear electrodynamics}}, \href{http://dx.doi.org/10.1016/j.aop.2015.08.028} {\emph{Ann. Phys.} (Amsterdam) {\bf 362}, (2015) 529}.

\bibitem{Balart.14.2014}
L.~Balart and E.~C. Vagenas, \emph{{Regular black hole metrics and the weak
  energy condition}},
  \href{http://dx.doi.org/10.1016/j.physletb.2014.01.024}{\emph{Phys. Lett. B}
  {\bf 730} (2014) 14--17}.

\bibitem{Balart.124045.2014}
L.~Balart and E.~C. Vagenas, \emph{{Regular black holes with a nonlinear
  electrodynamics source}},
  \href{http://dx.doi.org/10.1103/PhysRevD.90.124045}{\emph{Phys. Rev. D} {\bf
  90} (2014) 124045}.

\bibitem{Fan.124027.2016}
Z.-Y. Fan and X.~Wang, \emph{{Construction of regular black holes in general
  relativity}}, \href{http://dx.doi.org/10.1103/PhysRevD.94.124027}{\emph{Phys.
  Rev. D} {\bf 94} (2016) 124027}.

\bibitem{Lemos.124005.2011}
J.~P.~S. Lemos and V.~T. Zanchin, \emph{{Regular black holes: Electrically
  charged solutions, Reissner-Nordstr{\"o}m outside a de Sitter core}},
  \href{http://dx.doi.org/10.1103/PhysRevD.83.124005}{\emph{Phys. Rev. D} {\bf
  83} (2011) 124005}.

\bibitem{Ansoldi.0802.2008}
S.~Ansoldi, \emph{{Spherical black holes with regular center: a review of
  existing models}}, \href{https://arxiv.org/abs/0802.0330} {arXiv:0802.0330}.

\bibitem{Gott.243.1984}
J.~R. Gott and M.~Alpert, \emph{{General relativity in a (2 + 1)-dimensional
  space-time}}, \href{http://dx.doi.org/10.1007/BF00762539}{\emph{Gen.
  Relativ. Gravit.} {\bf 16} (1984) 243--247}.

\bibitem{Barrow.551.1986}
J.~D. Barrow, A.~B. Burd and D.~Lancaster, \emph{{Three-dimensional classical
  spacetimes}}, \href{http://dx.doi.org/10.1088/0264-9381/3/4/010}{\emph{Class.
  Quantum Grav.} {\bf 3} (1986) 551--567}.

\bibitem{Gott.1019.1986}
J.~R. Gott, J.~Z. Simon and M.~Alpert, \emph{{General relativity in a
  (2+1)-dimensional space-time: An electrically charged solution}},
  \href{http://dx.doi.org/10.1007/BF01090483}{\emph{Gen.
  Relativ. Gravit.} {\bf 18}
  (1986) 1019--1035}.

\bibitem{Giddings.751.1984}
S.~Giddings, J.~Abbott and K.~Kucha{\v{r}}, \emph{{Einstein's theory in a
  three-dimensional space-time}},
  \href{http://dx.doi.org/10.1007/BF00762914}{\emph{Gen.
  Relativ. Gravit.} {\bf 16}
  (1984) 751--775}.

\bibitem{Banados.1849.1992}
Ba{\~n}ados, Teitelboim and Zanelli, \emph{{Black hole in three-dimensional
  spacetime}}, \href{http://dx.doi.org/10.1103/PhysRevLett.69.1849}{\emph{Phys.
  Rev. Lett.} {\bf 69} (1992) 1849--1851}.

\bibitem{Banados.1506.1993}
M.~Ba{\~n}ados, M.~Henneaux, C.~Teitelboim and J.~Zanelli, \emph{{Geometry of
  the 2+1 black hole}},
  \href{http://dx.doi.org/10.1103/PhysRevD.48.1506}{\emph{Phys. Rev. D} {\bf
  48} (1993) 1506--1525}.

\bibitem{Chan.6385.1994}
K.~C.~K. Chan and R.~B. Mann, \emph{{Static charged black holes in
  (2+1)-dimensional dilaton gravity}},
  \href{http://dx.doi.org/10.1103/PhysRevD.50.6385}{\emph{Phys. Rev. D} {\bf
  50} (1994) 6385--6393}.

\bibitem{Zaslavskii.L33.1994}
O.~B. Zaslavskii, \emph{{Thermodynamics of 2+1 black holes}},
  \href{http://dx.doi.org/10.1088/0264-9381/11/2/003}{\emph{Class. Quantum
  Grav.} {\bf 11} (1994) L33--L38}.

\bibitem{Carlip.2853.1995}
S.~Carlip, \emph{{The (2 + 1)-dimensional black hole}},
  \href{http://dx.doi.org/10.1088/0264-9381/12/12/005}{\emph{Class. Quantum
  Grav.} {\bf 12} (1995) 2853--2879}.

\bibitem{Kamata.196.1995}
M.~Kamata and T.~Koikawa, \emph{{The electrically charged BTZ black hole with
  self (anti-self) dual Maxwell field}},
  \href{http://dx.doi.org/10.1016/0370-2693(95)00583-7}{\emph{Phys. Lett. B}
  {\bf 353} (1995) 196--200}.

\bibitem{Lemos.46.1995}
J.~Lemos, \emph{{Three dimensional black holes and cylindrical general
  relativity}},
  \href{http://dx.doi.org/10.1016/0370-2693(95)00533-Q}{\emph{Phys. Lett. B}
  {\bf 353} (1995) 46--51}.

\bibitem{Cataldo.2971.1996}
M.~Cataldo and P.~Salgado, \emph{{Static Einstein-Maxwell solutions in 2+1
  dimensions}}, \href{http://dx.doi.org/10.1103/PhysRevD.54.2971}{\emph{Phys.
  Rev. D} {\bf 54} (1996) 2971--2974}.

\bibitem{Clement.70.1996}
G.~Cl{\'e}ment, \emph{{Spinning charged BTZ black holes and self-dual
  particle-like solutions}},
  \href{http://dx.doi.org/10.1016/0370-2693(95)01464-0}{\emph{Phys. Lett. B}
  {\bf 367} (1996) 70--74}.

\bibitem{Hirschmann.5579.1996}
E.~W. Hirschmann and D.~L. Welch, \emph{{Magnetic solutions to 2+1 gravity}},
  \href{http://dx.doi.org/10.1103/PhysRevD.53.5579}{\emph{Phys. Rev. D} {\bf
  53} (1996) 5579--5582}.

\bibitem{Martinez.104013.2000}
C.~Mart{\'i}nez, C.~Teitelboim and J.~Zanelli, \emph{{Charged rotating black
  hole in three spacetime dimensions}},
  \href{http://dx.doi.org/10.1103/PhysRevD.61.104013}{\emph{Phys. Rev. D} {\bf
  61} (2000) 104013}.

\bibitem{Yamazaki.2001}
R. Yamazaki and D. Ida, \emph{{Black holes in three-dimensional Einstein-Born-Infeld-dilaton theory}},
  \href{http://dx.doi.org/10.1103/PhysRevD.64.024009}{\emph{Phys. Rev. D} {\bf
  64} (2001) 024009}.

\bibitem{Mazharimousavi.124021.2011}
S.~H. Mazharimousavi, O.~Gurtug, M.~Halilsoy and O.~Unver, \emph{{2+1
  dimensional magnetically charged solutions in Einstein-power-Maxwell
  theory}}, \href{http://dx.doi.org/10.1103/PhysRevD.84.124021}{\emph{Phys.
  Rev. D} {\bf 84} (2011) 124021}.

\bibitem{Gurtug.104004.2012}
O.~Gurtug, S.~H. Mazharimousavi and M.~Halilsoy, \emph{{2+1 -dimensional
  electrically charged black holes in Einstein-power-Maxwell theory}},
  \href{http://dx.doi.org/10.1103/PhysRevD.85.104004}{\emph{Phys. Rev. D} {\bf
  85} (2012) 104004}.

\bibitem{Xu.124008.2013}
W.~Xu and L.~Zhao, \emph{{Charged black hole with a scalar hair in ( 2+1 )
  dimensions}}, \href{http://dx.doi.org/10.1103/PhysRevD.87.124008}{\emph{Phys.
  Rev. D} {\bf 87} (2013) 124008}.

\bibitem{Frassino.124069.2015}
A.~M. Frassino, R.~B. Mann and J.~R. Mureika, \emph{{Lower-dimensional black
  hole chemistry}},
  \href{http://dx.doi.org/10.1103/PhysRevD.92.124069}{\emph{Phys. Rev. D} {\bf
  92} (2015) 124069}.

\bibitem{Trodden.483.1993}
M.~Trodden, V.~F. Mukhanov and R.~H. Brandenberger, \emph{{A nonsingular two
  dimensional black hole}},
  \href{http://dx.doi.org/10.1016/0370-2693(93)91032-I}{\emph{Phys. Lett. B}
  {\bf 316} (1993) 483--487}.

\bibitem{Cataldo.084003.2000}
M.~Cataldo and A.~Garc{\'i}a, \emph{{Regular (2+1)-dimensional black holes
  within nonlinear electrodynamics}},
  \href{http://dx.doi.org/10.1103/PhysRevD.61.084003}{\emph{Phys. Rev. D} {\bf
  61} (2000) 084003}.

\bibitem{Myung.405.2009}
Y.~S. Myung and M.~Yoon, \emph{{Regular black hole in three dimensions}},
  \href{http://dx.doi.org/10.1140/epjc/s10052-009-1036-9}{\emph{Eur. Phys. J.
  C} {\bf 62} (2009) 405--411}.

\bibitem{Mazharimousavi.893.2012}
S.~H. Mazharimousavi, M.~Halilsoy and T.~Tahamtan, \emph{{Regular charged black
  hole construction in (2+1) dimensions}},
  \href{http://dx.doi.org/10.1016/j.physleta.2012.01.001}{\emph{Phys. Lett. A}
  {\bf 376} (2012) 893--898}.

\bibitem{Hayward.031103.2006}
S.~A. Hayward, \emph{{Formation and evaporation of nonsingular black holes}},
  \href{http://dx.doi.org/10.1103/PhysRevLett.96.031103}{\emph{Phys. Rev.
  Lett.} {\bf 96} (2006) 031103}.


\bibitem{Miskovic.024011.2011}
O.~Mi{\v{s}}kovi{\'c} and R.~Olea, \emph{{Conserved charges for black holes in
  Einstein-Gauss-Bonnet gravity coupled to nonlinear electrodynamics in AdS
  space}}, \href{http://dx.doi.org/10.1103/PhysRevD.83.024011}{\emph{Phys. Rev.
  D} {\bf 83} (2011) 024011}.

\bibitem{Breton.643.2005}
N.~Bret{\'o}n, \emph{{Smarr's formula for black holes with non-linear
  electrodynamics}},
  \href{http://dx.doi.org/10.1007/s10714-005-0051-x}{\emph{Gen. Relativ.
  Gravit.} {\bf 37} (2005) 643--650}.

\bibitem{Balart.280.2010}
L.~Balart, \emph{{Quasilocal energy, Komar charge and horizon for regular black
  holes}}, \href{http://dx.doi.org/10.1016/j.physletb.2010.03.056}{\emph{Phys.
  Lett. B} {\bf 687} (2010) 280--285}.

\bibitem{Ma.245014.2014}
M.-S. Ma and R.~Zhao, \emph{{Corrected form of the first law of thermodynamics
  for regular black holes}},
  \href{http://dx.doi.org/10.1088/0264-9381/31/24/245014}{\emph{Class. Quantum
  Grav.} {\bf 31} (2014) 245014}.
  
\bibitem{Ma.CPL}
M.-S. Ma and R.~Zhao, \emph{{Notes on Phase Transition of Nonsingular Black Hole}}, \href{http://dx.doi.org/10.1088/0256-307X/32/3/030401}{\emph{Chin.Phys.Lett.} {\bf 32} (2015) 030401}.

\bibitem{Cvetic.2011}
M. Cveti\v{c}, G. W. Gibbons, D. Kubiz\v{n}\'{a}k, and C. N. Pope, \emph{{Black hole enthalpy and an entropy inequality for the thermodynamic volume}}, \href{http://dx.doi.org/10.1103/PhysRevD.84.024037}{\emph{Phys. Rev.
  D} {\bf 84} (2011) 024037}.








\end{thebibliography}

\end{document}